\documentclass[12pt,a4paper]{article}
\usepackage[]{amssymb}
\begin{document}
\title{%
\begin{flushright}
\normalsize
ULB-TH-0525\\
\end{flushright}
\protect\vspace{5mm}
Higgs boson with a single generation in the bulk.
}
\author{M.V.~Libanov$^{1,2}$ and
E.Y.~Nugaev$^{1}$\\
\small\em
$^{1}$~Institute for Nuclear Research of the Russian Academy of
Sciences,\\
\small\em
60th October Anniversary Prospect 7a, 117312, Moscow, Russia;\\
\small\em
$^{2}$~Service de Physique Th\'{e}orique, CP 225,\\
\small\em
  Universit\'{e} Libre de Bruxelles, B--1050, Brussels, Belgium;\\
\small\em
}
\date{}
\maketitle
\vspace{-12mm}
\begin{abstract}
We study Higgs boson properties in a model where three fermionic
families of the Standard Model arise from a single generation in (5+1)
dimensions. We demonstrate that, in spite of a non-trivial background,
properties of the four-dimensional Higgs particle are almost
indistinguishable from those in the Standard Model. We also argue
that it is more natural to have a light Higgs boson in  this model.
\end{abstract}
\newpage
\section{Introduction and summary}

Large extra dimensions is a fruitful idea for constructing new models of
high energy physics (see e.g.\ Ref.~\cite{LED}; for a review see
Ref.~\cite{RuUFN} and for earlier work see Refs.~\cite{Anton}). In
particular, in the framework of large extra dimensions \cite{LED}, the
models have been suggested \cite{LT, FLT} and studied \cite{FLTnu, LN}
where three generations of the Standard Model (SM) fermions appear as
three zero modes localized in the four-dimensional core of a defect with
topological number three. When both fermions and expectation value of
the Higgs field are localized on the brane, the
corresponding overlaps may result in a hierarchical
pattern of fermion masses and mixings \cite{Schmaltz}. This occurs
naturally in the models under discussion \cite{LT}. To incorporate
four-dimensional gauge fields, a compactified version of the model has
been developed \cite{FLNT} (other possible mechanisms  gauge fields
localization are discussed in, e.g., Ref.~\cite{gaugelocaliz}). There,
due to the specific Yukawa coupling fermions  are localized in the core of a
(5+1)-dimensional vortex with winding number one, and two extra
dimensions which form a sphere with radius $R$, are accessible for
(non-localized) gauge bosons. The zero modes of the gauge bosons are
homogeneous over the sphere, while higher modes have
non-trivial profiles, and hence different overlaps with fermionic zero
modes. Since in this model three four-dimensional families originate from
a single six-dimensional generation, flavour violation processes appear in
the effective four-dimensional theory~\cite{FLNTflavour}. Naively, one
could expect that amplitudes of such processes are suppressed only by
the mass of the first flavour violating excitation of, say, $Z$-boson. The
mass of that $Z'$-boson is proportional to $1/R$. So, having at hand
experimental bounds on probabilities of flavor violating processes one
expects to find a bound on the size of extra dimensions\footnote{More
precisely, a bound on the scale of gauge fields localization if one uses
a more elaborate mechanism of the localization.}. It turns out, however,
that an approximate rotation symmetry in extra dimensions results in
additional suppression of flavour changing neutral currents (FCNC) which
also violate generation number. As a result, the strongest bound on $R$
arises from the branching ratio of the decay $K_L\to \mu e$. It has been
shown in Ref.~\cite{FLNTflavour} that the bound is $1/R \gtrsim 60$ TeV.
In general, one may expect that more elaborate mechanisms of the gauge
fields localization may reduce this bound and may allow for  $1/R$ of
order several TeV. Possible signatures of new FCNC bosons ($Z',\gamma '$)
for collider physics were discussed in Ref.~\cite{collider}.

Here, we study the Higgs boson which arises in our model as the first
excitation of the background Higgs field. Due to the nontrivial background
of the Higgs field, that has non-zero value near the core and tends to zero
outside the core, one may in principle expect some new properties of the
Higgs boson as compared to the SM. In particular, one may expect that the
Higgs boson could be heavier than the SM one, or even could be
delocalized and, being produced in a collider, disappears from our
brane and  propagates freely into extra dimensions. In any case, one may
expect that the relationships between the mass, cubic and quartic couplings
are not the same as in the SM. All these peculiarities could be
very interesting from a phenomenological point of view. In this paper we
clarify these issues. Starting from the
hierarchical fermionic mass pattern~\cite{FLNT}, masses of the usual $Z$
and $W$, and the bound on the size of the extra
dimensions~\cite{FLNTflavour} we will show that the value of the Higgs
background in the core should be very small compared to typical scales of
the model. This requires some tuning of parameters of the model,
and implies that the Higgs boson in our model is almost indistinguishable
from the Higgs boson of the SM. We will also argue that it is more
natural to have a light  Higgs boson in our model, with mass not much
exceeding 100 GeV.

\section{Setup}
\label{sec:2}

In this paper we consider the case of sphere $S^2$ of radius $R$ as
compact extra dimensions as a concrete example. Introducing  spherical
coordinates  $(\theta ,\varphi) $ with the north pole  ($\theta=0$) in the
centre of the core (the latter occupies a  small region of $S^2$), the
distance from the core to a point is equal to $R \theta$. Using this
length as a variable in the calculations, our results can be
straightforwardly transferred to models with  flat extra dimensions.  In
this paper we restrict ourselves to the case of two scalar fields. The
first one, $\Phi $, is charged only under $U_g(1)$ gauge group with charge
$+1$ and, together with the $U_g(1)$ gauge field $A_A$, forms the
Abrikosov-Nielsen-Olesen vortex\footnote{Our notations coincide with those
used in Refs.~\cite{LT, FLT, FLNT}. In particular, six-dimensional
coordinates  are labeled by capital Latin indices $A,B=0,\dots,5$.
Four--dimensional coordinates are labeled by Greek indices $\mu,\nu
=0,\dots,3$. The metric is
$g_{AB}=\mbox{diag}(1,-1,-1,-1,-R^2,-R^2\sin^2\theta )$.}. The second
field, $H$, is charged under $U_g(1)$ with charge $-1$ and  carries
quantum numbers of the Higgs doublet under $SU(2)\times U(1)_Y$ SM gauge
group. For some set of parameters, this field develops a non-trivial
profile in  extra dimensions and gives rise to electroweak symmetry
breaking. The fermionic mass matrix and mixings are reproduced once the third
scalar field $X$ is introduced (see Ref.~\cite{FLT} for details). The
field $X$ is irrelevant for the present study, and we do not write it in
what follows.

The potential term in the Lagrangian which gives rise to  non-trivial
profiles of the scalar fields has the following form
\begin{equation}
V_s= {\lambda\over 2}\left( |\Phi|^2-v^2\right)^2+ {\kappa\over 2} \left(
|H|^2-\mu^2\right)^2+h^2|H|^2|\Phi|^2
\label{ScalarPotential}
\end{equation}
A coupling with the $U_g(1)$-gauge vector field $A_A$ is introduced in the
usual way
\[
\partial_A\to\partial_A\pm ieA_A,
\]
where upper and lower signs correspond to the Higgs and $\Phi $
fields; $e$ is a new dimensional parameter of the theory. To reproduce the
hierarchical fermionic mass pattern one  chooses  small value of
$e$~\cite{FLNT},
\[
e\ll \sqrt \lambda.
\]
In this case, vortices with winding number one are stable.

The set of static equations for the background has the following form
(prime denotes the derivative with respect to $\theta $),
\begin{eqnarray}
&&\frac{1}{\sin \theta }(\sin \theta F')'-\frac{1}{\sin^2\theta
}F(A-1)^2-\lambda R^2F(F^2-v^2)-R^2h^2FH^2=0\;,\nonumber\\
&&\frac{1}{\sin \theta }(\sin \theta H')'-\frac{1}{\sin^2\theta
}HA^2-\kappa  R^2H(H^2-\mu ^2)-R^2h^2HF^2=0\;,\nonumber\\
&&\frac{1}{\sin \theta }\left(\frac{A'}{\sin\theta }
\right)'-\frac{2R^2e^2}{\sin^2\theta }F^2(A-1)-
\frac{2R^2e^2}{\sin^2\theta }H^2A=0\;,
\label{Eq/Pg3/1:HIGGS}
\end{eqnarray}
with the boundary conditions (see~\cite{FLNT} for details)
\begin{eqnarray}
&&F(\theta )=0\;,\ H'(\theta )=0\;,\ A(\theta )=0\;,\ \ \mbox{at }\theta
=0\;,\nonumber\\
 &&F'(\theta )=0\;,\ H(\theta )=0\;,\ A(\theta )=1\;,\ \ \mbox{at
}\theta=\pi
 \label{Eqn/Pg3/1:HIGGS}
\end{eqnarray}
Here we use the anzatz for Abrikosov-Nielsen-Olesen vortex with winding
number one in the spherical coordinates ($\alpha =1,2$ is the $SU(2)$
index):
\[
 A_\varphi =\frac{1}{e}A(\theta )\;,\ \ \ A_\theta =0\;,\ \ \ \Phi
 =F(\theta )\mathrm{ e}^{i\varphi }\;,\ \ \ H_\alpha =\delta _{\alpha
2}H(\theta )
 \]
The set of the equations (\ref{Eq/Pg3/1:HIGGS}),
(\ref{Eqn/Pg3/1:HIGGS}) admits a non-trivial solution. The existence of
solutions for $\Phi $ and for the gauge field is guarantied by
topology~\cite{FLNT}; these fields form the usual vortex. Making use of
the arguments of Ref.~\cite{Witten}, one can show that with $h^2v^2>\kappa
\mu ^2$, there is a stable non-trivial solution for $H(\theta )$ which is
non-zero inside the core of the vortex. It is worth noting that the
boundary conditions (\ref{Eqn/Pg3/1:HIGGS}) do not fix the amplitude of
$H(\theta )$ at the origin. Naively, one can expect that this amplitude is
of order $\mu $. However, as we demonstrate below, to make this setup
phenomenologically viable, this amplitude should be very small compared to
$\mu $.

First, one notes that the size of the region with non-vacuum
field $\Phi$, $R\theta _\Phi \simeq 1/\sqrt{\lambda }v$ coincides with
that for the Higgs field $R\theta _H$, that is, with the scale where $H$
is appreciably non-zero (see  Ref.~\cite{LN} for details and the next
section for discussions). Numerical simulations show that
\begin{equation}
\theta _H=\frac{a}{R\sqrt{\lambda }v }
\label{Eq/Pg4/1B:HIGGS}
\end{equation}
where $a\simeq 4.5$ and very weakly depends on $\lambda ,v$ and $R$.

Second, to reproduce the hierarchy of the fermionic masses one
requires that $\theta _H /\theta _A\sim 0.1$~\cite{FLNT} where $R\theta
_A\sim 1/(ev)$ is the size of the gauge core. Thus, in any case
\begin{equation}
\theta _H \lesssim 0.1
\label{Eq/Pg4/1A:HIGGS}
\end{equation}
Third, as we mentioned above $H$ is the only scalar charged under the
electroweak group. The non-trivial solution for the Higgs-vortex system
breaks the electroweak symmetry down, and gives rise to non-zero values of
four-dimensional masses of $SU(2)$ gauge bosons. One introduces the
effective Higgs expectation value
\begin{equation}
2\pi R^2\int \limits_{0}^{\pi }\!d\theta \sin\theta H^2(\theta )\equiv
\frac{V_\mathit{ SM}^2}{2}
\label{Eq/Pg4/2:HIGGS}
\end{equation}
Then the gauge boson masses  can be expressed through $V_\mathit{ SM}$ in
the same way as in the SM, that is, for instance, $m_W=gV_\mathit{
SM}/2$ where $g$ is the usual four-dimensional $SU(2)$ gauge
coupling~\cite{FLNTflavour}. In other words, $V_\mathit{ SM}\simeq 250$
GeV, the standard Higgs vacuum expectation value.

By making use  of a step-like approximation for $H(\theta )$ with
amplitude $H(0)$ and width $\theta _H$, Eqs.~(\ref{Eq/Pg4/1B:HIGGS}),
(\ref{Eq/Pg4/1A:HIGGS}), one estimates the integral in
(\ref{Eq/Pg4/2:HIGGS}). With $V_\mathit{ SM}\simeq 250$ GeV one then finds
\begin{equation}
H(0)=\frac{1}{\sqrt{2\pi}a^2}(V_\mathit{ SM}\cdot R\theta _H) \lambda
v^2\sim 10^{-5} \lambda v^2
\label{Eq/Pg5/1:HIGGS}
\end{equation}
where we have used the bound $1/R>60$ TeV from ~\cite{FLNTflavour} and
also used Eq.(\ref{Eq/Pg4/1A:HIGGS}).

Next, let us choose
\begin{equation}
\lambda v^2\sim \kappa \mu ^2
\label{Eq/Pg5/1A:HIGGS}
\end{equation}
This is  quite natural choice; we will have to say more about it later. Let
us note also that $\lambda v$ as well as $\kappa \mu $ cannot be much
larger than one. Indeed, the dimensionless combination $\lambda v/(4\pi
)^3$ plays a role of the coupling constant in six-dimensional theory. So,
to avoid  strong coupling regime in the scalar theory
(\ref{ScalarPotential}) at length scales of interest, we assume that
\begin{equation}
\lambda v\sim \kappa \mu \sim 1
\label{Eq/Pg5/2:HIGGS}
\end{equation}
Therefore, making use of Eqs.~(\ref{Eq/Pg5/1A:HIGGS}) and
(\ref{Eq/Pg5/2:HIGGS}) one finds from (\ref{Eq/Pg5/1:HIGGS})
\begin{equation}
H(0)\simeq 10^{-5}\mu
\label{Eq/Pg5/3:HIGGS}
\end{equation}
in contrast to the naive expectation. This means, in particular, that one
has to fine tune parameters of the theory. In the next section we will see
how it can be done.

\section{Fine tuning}
\label{Section/Pg5/1:HIGGS/Fine tuning}
To understand fine tuning of
parameters leading to a solution satisfying
(\ref{Eq/Pg5/3:HIGGS}) let us study the spectrum of small
perturbations about the solution $H_{\alpha} =\delta _{2\alpha }H(\theta
)$. In general, there are four real excitations. Three of them (two
excitations of $H_1$ and imaginary part of the perturbation of $H_2$)
are zero modes corresponding to spontaneous breaking of  $SU(2)$
symmetry. From the four-dimensional point of view these excitations are the
usual $SU(2)$ Goldstone bosons which are eaten up to give  masses  to
the gauge bosons. So, in what follows we will not consider these modes. Let
us concentrate on the real part of the perturbations about $H_2$
(here $p_\mu $ is  four-dimensional momentum)
\[
H_2=H(\theta )+\chi \cdot\mathrm{ e}^{-ip_\mu x^\mu }
\]
We will consider only the lowest $s$-wave excitation, so $\chi =\chi
(\theta )$; it corresponds to the wave function of the Higgs particle  in
extra dimensions.

The equation of motion for $\chi $ is
\begin{equation}
-\chi ''- {\rm ctg} \theta \chi '+U(\theta )\chi =R^2\omega^2 \chi
\label{Eq/Pg1/2:sh}
\end{equation}
where
\begin{equation}
U(\theta )=\frac{A^2}{\sin \theta ^2} +\kappa
R^2(3H^2-\mu ^2)+h^2R^2F^2
\label{Eq/Pg6/1:HIGGS/POT}
\end{equation}
is effective quantum mechanical potential; $\omega ^2\equiv p_\mu ^2$ is
four-dimensional mass of the Higgs particle at $\omega ^2\geq 0$.
The wave function $\chi (\theta )$ obeys the following
normalization condition,
\begin{equation}
2\pi R^2\int \limits_{0}^{\pi }\!d\theta \sin\theta \chi ^2=\frac{1}{2}
\label{Eq/Pg6/2:HIGGS}
\end{equation}

Let us briefly repeat the Witten's argument~\cite{Witten} showing that
there exists  non-trivial $H(\theta )$ if $h^2v^2>\kappa
\mu ^2$. One considers a trivial solution $H(\theta )=0$ of
(\ref{Eq/Pg3/1:HIGGS}) and perturbations about it to show that there is a
mode with $\omega ^2<0$.
At $\theta >\theta _\Phi $ $F(\theta )$
tends to its vacuum expectation value $v$. If $\theta _\Phi $ is small
enough then the first term in the r.h.s. of (\ref{Eq/Pg6/1:HIGGS/POT}) is
important in the vicinity of the south pole ($\theta =\pi $) only (note
that $A(\theta )\sim (evR\theta) ^2$ at $\theta \to 0$). So, one
neglects this term for a while\footnote{This term appears  because $H$ is
charged under $U_g(1)$. So, if $H$ is a singlet under $U_g(1)$ or one
considers flat extra dimensions, this term plays no role at all.}. Thus,
for $H=0$, the potential (\ref{Eq/Pg6/1:HIGGS/POT}) has the following
behaviour. It starts from $-R^2\kappa \mu ^2$ at the origin and tends to
$R^2(h^2v^2-\kappa \mu ^2)$ at large $\theta $. In two dimensions such
potentials always possess bound states. If $h^2v^2=\kappa \mu ^2$, then
the corresponding eigenvalue is certainly negative and the trivial
solution is unstable. Moreover, if the condition
(\ref{Eq/Pg5/1A:HIGGS}) holds then $\omega ^2$ is not exponentially
small: the range of the potential is of order $\theta _\Phi \sim
1/\sqrt{\lambda v^2R^2}$, and the dimensionless product
(range)$^2\times$(strength of potential)$=(\kappa \mu ^2)/(\lambda v^2)$
is of order one. So, by continuity, there is also a bound state solution
with negative $\omega ^2$ at $h^2v^2>\kappa \mu ^2$, and, therefore, a
stable non-trivial solution should exist. This explains the condition
(\ref{Eq/Pg5/1A:HIGGS}).

On the other hand, if the combination $h^2v^2$ increases while $\kappa
\mu ^2$  is left intact, then the negative mode about the trivial solution
may disappear. This means in turn, that the non-trivial solution may
become unstable. That happens when the non-trivial solution, whose
behaviour at large $\theta $ is $\exp(-\theta R\sqrt{h^2v^2-\kappa \mu
^2})$, becomes narrow compared to $F(\theta )$. In this case the potential
(\ref{Eq/Pg6/1:HIGGS/POT}) has  a negative minimum at non-zero $\theta \ll
\theta _\Phi $ which leads to a negative mode about the non-trivial
solution\footnote{It is worth noting that the solutions $F(\theta )$
to Eq.~(\ref{Eq/Pg3/1:HIGGS}) are different for $H(\theta )=0$ and
$H(\theta )\neq 0$ (especially when $h^2v^2$ is large). So, there is no
contradiction between the existence of a positive modes only for $H=0$ and
a negative one for the non-trivial solution.}. This fact explains why one
chooses $\theta _H\simeq \theta _\Phi $.

All these arguments allow to fine tune parameters and to construct  a
non-trivial solution with $H(0)\simeq 10^{-5}\mu $. A simplest
way to construct the solution is to tune the parameter $\kappa $. Indeed,
let us fix $h$, $v\simeq \mu $ and choose $\kappa _B$ satisfying
$h^2v^2=\kappa _B\mu ^2$. In this case there certainly exists a negative
mode $\chi _B(\theta )$ with an eigenvalue $\omega _B^2<0$ for the trivial
solution $H(\theta )=0$. Let us choose a new value for $\kappa$ equal to
$\kappa _0$ where
\[
\kappa _0=\kappa _B-\frac{|\omega _B^2|}{\mu ^2}
\]
Numerically, we find $\kappa _0\simeq 0.8 \kappa _B$. Note that in the
case $H(\theta )=0$  shifts of $\kappa $ are the overall shifts of the
potential (\ref{Eq/Pg6/1:HIGGS/POT}) and, hence, do not change the
corresponding wave function. Thus, with this new $\kappa _0$ the function
$\chi _B$ becomes exactly the zero mode of Eq.~(\ref{Eq/Pg1/2:sh}). Let us
now shift $\kappa _0$ to
\begin{equation}
\kappa _T =\kappa _0(1+\alpha )
\label{Eq/Pg7/1B:HIGGS}
\end{equation}
where $\alpha $ is a small positive number. Then again the trivial
solution becomes unstable: its negative mode $\chi _B$ corresponds to the
eigenvalue
\begin{equation}
\omega ^2=-\alpha \kappa _0\mu ^2=-\alpha \kappa _T\mu ^2+\mathcal{
O}(\alpha ^2)
\label{Eq/Pg7/1A:HIGGS}
\end{equation}
Note that  $\chi _B$ determines the direction in the Hilbert space from
the unstable solution $H(\theta )=0$ to the stable non-trivial
configuration. In other words, the non-trivial stable solution can be
represented as follows
\begin{equation}
H(\theta )=\beta \chi _B+\chi _\bot
\label{Eq/Pg7/1:HIGGS}
\end{equation}
where $\beta $ is a small number,  $\chi _\bot$ is
orthogonal to $\chi _B$:
\begin{equation}
R^2\int \limits_{0}^{2\pi }d\varphi \int \limits_{0}^{\pi }\!d\theta
\sin\theta (\chi _B\cdot\chi _\bot)=0
\label{Eq/Pg8/1:HIGGS}
\end{equation}
and, furthermore, $\chi _\bot$ is suppressed compared to $\chi _B$ at small
$\alpha $. Note that $\chi _\bot$ remains in the space orthogonal to
$\chi_B$ upon the action of the operator in the l.h.s. of Eq.
(\ref{Eq/Pg1/2:sh}).

To find $\beta $ one  substitutes the expansion (\ref{Eq/Pg7/1:HIGGS})
into Eq. (\ref{Eq/Pg3/1:HIGGS}) with $\kappa =\kappa _T$ (Eq.
(\ref{Eq/Pg7/1B:HIGGS})), uses Eq. (\ref{Eq/Pg1/2:sh}) for $\chi _B$ with
$\omega ^2$ given by (\ref{Eq/Pg7/1A:HIGGS}), calculates the inner product
with $\chi _B$ by making use of Eqs. (\ref{Eq/Pg6/2:HIGGS}),
(\ref{Eq/Pg8/1:HIGGS}), and neglects  terms like $\langle \chi _B^3, \chi
_\bot\rangle $, which are subdominant since $\chi_\bot$ is suppressed. In
this way one obtains the following equation
\begin{equation}
\alpha \mu ^2=\beta ^2\cdot \left(4\pi R^2\int \limits_{0}^{\pi }
\!d\theta \sin\theta \chi ^4_B \right)
\label{Eq/Pg8/2:HIGGS}
\end{equation}
Due to the normalization condition (\ref{Eq/Pg6/2:HIGGS}) the expression
in the parenthesis in the r.h.s. of (\ref{Eq/Pg8/2:HIGGS}) is of order
$R^{-2}$. Therefore, we have
\[
\beta \simeq R\mu \cdot\sqrt{\alpha }
\]
and
\[
H(0)\simeq \mu \sqrt{\alpha }\cdot\left(R\chi_B (0)   \right)
\]
Again, $R\cdot\chi _B(0)$ is some finite number, and to have $H(0)$ of
order $10^{-5}\mu $ one should choose $\alpha \sim 10^{-10}$.

It is worth noting that the method described is useful in numerical
analysis of the boundary value problem (\ref{Eq/Pg3/1:HIGGS}),
(\ref{Eqn/Pg3/1:HIGGS}). A common tool in such simulations is the
relaxation method~\cite{Num}. This method requires, however, an initial
guess for the solution. This initial guess is naturally given by Eq.
(\ref{Eq/Pg7/1:HIGGS}) with $\chi _\bot=0$ and $\beta $ determined by
(\ref{Eq/Pg8/2:HIGGS}). We have solved the boundary value problem
(\ref{Eq/Pg3/1:HIGGS}), (\ref{Eqn/Pg3/1:HIGGS}) in this way and present
the results of our calculations in appropriate places.

\section{Higgs particle in four-dimensional theory}

As we have already mentioned, $\chi (\theta )$ is the transverse wave
function  of the standard Higgs particle. Let us
study now properties of the Higgs boson.
First, we note that $\chi (\theta )$ is the solution
of (\ref{Eq/Pg1/2:sh}) with non-trivial background $H(\theta )$.
Nevertheless, since the amplitude of $H(\theta )$ is very small, its
contribution to (\ref{Eq/Pg1/2:sh}) is negligible, and, hence, $\chi $
coincides with $\chi _B$. So, in what follows we will use the following
approximations for the background field
\begin{equation}
H(\theta )=\beta \chi _B(\theta )
\label{Eq/Pg9/1:HIGGS}
\end{equation}
and $\chi _B$ is still the solution of (\ref{Eq/Pg1/2:sh}) with $H(\theta
)=0$.

Since $\chi _B$ is a bound state, the Higgs boson cannot disappear from
our world. However, due to a non-trivial profile of $\chi_B $ one might
expect that the relationship between the four-dimensional effective
coupling constants of the Higgs boson could be different from the usual
one. We are going to show now that this is not the case.

To begin with, let us express $\beta $ in Eq. (\ref{Eq/Pg9/1:HIGGS})
through $V_\mathit{ SM}$. Making use of the definition
(\ref{Eq/Pg4/2:HIGGS}), the normalization condition (\ref{Eq/Pg6/2:HIGGS})
and Eq. (\ref{Eq/Pg9/1:HIGGS}) one immediately finds
\begin{equation}
\beta =V_\mathit{ SM}
\label{Eq/Pg9/2:HIGGS}
\end{equation}

Let us now find the SM quartic coupling of the Higgs boson. This coupling
arises from $|H|^4$ term in (\ref{ScalarPotential}) after integration over
extra dimensions. So, we have\footnote{Hereafter, we assume that
$\kappa $ is properly fine tuned as described in the Section
\ref{Section/Pg5/1:HIGGS/Fine tuning} and  skip the subscript $T$.}
\begin{equation}
\lambda _{\mathit{ SM}}=4\pi R^2\kappa \int \limits_{0}^{\pi }\!d\theta
\sin\theta \chi_B ^4
\label{Eq/Pg9/3:HIGGS}
\end{equation}
This integral can be easily estimated. To this end one can use a linear
approximation for $\chi _B$ at $\theta <\theta _H$:
\begin{equation}
\chi _B=\frac{\sqrt{3}}{R\theta _H}\cdot\left(1-\frac{\theta }{\theta _H}
\right)
\label{Eq/Pg10/1:HIGGS}
\end{equation}
and $\chi _B=0$ at $\theta >\theta _H$ where $\theta _H$ is given by
(\ref{Eq/Pg4/1B:HIGGS}). The constant in the r.h.s. of
(\ref{Eq/Pg10/1:HIGGS}) is determined from the normalization
condition (\ref{Eq/Pg6/2:HIGGS}). In this way one obtains
\begin{equation}
\lambda_\mathit{ SM}\simeq \frac{6}{5a^2\pi } \frac{\kappa }{\lambda
}(\lambda v)^2
\label{Eq/Pg10/2:HIGGS}
\end{equation}
For $\lambda =2.4/v$, $\kappa =2.1/v$, $\mu =v$, $a=4.5$ this formula
gives $\lambda _{\mathit{ SM}}=0.095$ while the numerical solution gives
$\lambda _\mathit{SM}=0.112$ and shows that the result indeed does not
depend on $R$.

The cubic coupling is determined by another integral. Using
(\ref{Eq/Pg9/1:HIGGS}), (\ref{Eq/Pg9/2:HIGGS}) and (\ref{Eq/Pg9/3:HIGGS}),
we have
\[
g_{3,SM}=4\pi \kappa R^2\int \limits_{0}^{\pi }\!d\theta \sin\theta
\chi_B ^3H =\lambda _\mathit{ SM}v_\mathit{ SM}
\]
that is, precisely the same relation which appears in the SM. This happens,
of course, due to the relation (\ref{Eq/Pg9/1:HIGGS}).

To find the Higgs boson mass which is the lowest eigenvalue of Eq.
(\ref{Eq/Pg1/2:sh}) in the non-trivial background (\ref{Eq/Pg9/1:HIGGS}),
one makes use of the standard quantum mechanical perturbation theory,
considering the term $3\kappa R^2H^2$ as perturbation. Since $\chi _B$ is
the exact solution of (\ref{Eq/Pg1/2:sh}) with $\omega ^2$ given by
(\ref{Eq/Pg7/1A:HIGGS}), one has
\begin{equation}
m_H^2=\omega ^2 +3\kappa \cdot 4\pi R^2\int \limits_{0}^{\pi }\!d\theta
\sin\theta H^2\chi _B^2
\label{Eq/Pg10/3:HIGGS}
\end{equation}
The value of $\omega ^2$ can be expressed through $V_\mathit{ SM}$ and
$\lambda _\mathit{ SM}$ by means of (\ref{Eq/Pg8/2:HIGGS}),
(\ref{Eq/Pg9/2:HIGGS}), (\ref{Eq/Pg9/3:HIGGS}), while the integral in the
r.h.s. of (\ref{Eq/Pg10/3:HIGGS}) can be calculated by making use of
(\ref{Eq/Pg9/1:HIGGS}) and (\ref{Eq/Pg9/3:HIGGS}). In this way one finds
\[
m_H^2=2\lambda _\mathit{ SM}V_\mathit{ SM}^2
\]
that is again  the conventional SM expression for the Higgs boson mass. Using
the approximate value for $\lambda _\mathit{ SM}$ (\ref{Eq/Pg10/2:HIGGS})
one has an estimate
\[
m_H\simeq\sqrt{\frac{\kappa }{\lambda }}(\lambda v)\cdot 50 \mbox{GeV}
\]
Note that $\lambda v$ cannot be much  larger than 1 (see the
discussion in  Section \ref{sec:2}). It is interesting to note also, that
the ratio $\kappa /\lambda $ could not be large. Actually, at fixed
$\kappa \mu $ and $\lambda v$ the ratio $\lambda /\kappa $ determines the
strength of the potential in (\ref{Eq/Pg1/2:sh}) (see the discussion in
the Section \ref{Section/Pg5/1:HIGGS/Fine tuning}). So, if $\kappa /\lambda
$ is large enough the potential becomes weak and, neglecting the first
terms in (\ref{Eq/Pg6/1:HIGGS/POT}), the negative level about the trivial
solution becomes exponentially small. Rough estimate gives
\begin{equation}
|R\omega |\sim \frac{1}{\theta ^2_H }\exp\left[ -\frac{6}{a^2}\frac{\kappa
}{\lambda }\frac{(\lambda v)^2}{(\kappa \mu)^2 }\right]
\label{Eq/Pg11/2:HIGGS}
\end{equation}
So, at large enough $\kappa /\lambda $ the first repulsive term in
(\ref{Eq/Pg6/1:HIGGS/POT}) becomes relevant and the lowest
level is in fact positive. That is, the trivial solution is
stable. Assuming that $(\lambda v)/(\kappa \mu )\simeq 1$ one finds from
(\ref{Eq/Pg11/2:HIGGS}) that $\kappa /\lambda <4$. This in turn leads to
$m_H\simeq (\lambda v)\cdot 100$ GeV, which is close to the
experimental bound $m_H>114$ GeV. Thus, we see that it is more natural to
have the light Higgs boson in  our model.

To conclude, we stress that although we have considered the case
$1/R>60$ TeV, our results are almost insensitive to the choice of $R$
which is basically the scale of localization of the
SM gauge fields if one uses a more elaborate mechanism of the
localization. Indeed, our results are based on the fact that the amplitude
of the background $H$ is small compared to $\mu $. This fact, in turn,
follows from Eq.~(\ref{Eq/Pg5/1:HIGGS}): the smallness of $H(0)$ is
determined by the two small parameters $V_\mathit{ SM}\cdot R$ and $\theta
_H$. While the first parameter $V_\mathit{ SM} \cdot R$ could be larger,
 $\theta _H\sim 0.1$ is fixed by the hierarchy of the fermionic
masses~\cite{FLNT}. Thus, if due to some mechanism $R$ would turn out to
 be larger ($V_\mathit{ SM}\cdot R\sim 1$) then Eq.~(\ref{Eq/Pg5/3:HIGGS})
would give $H(0)\sim 10^{-3}\mu $, that is, again $H$ would be small
compared to $\mu $ and our results would remain intact.

\section*{Acknowledgments} The authors are indebted to S.~Dubovsky,
J.-M.~Frere, V.~Rubakov, and S.~Troitsky for helpful comments. M.L. thanks
Service de Physique Th\'{e}orique, Universit\'{e} Libre de Bruxelles,
where this work has been completed, for hospitality. This work has been
supported in part by RFBR grant 05-02-17363-a, by the Grants of the
President of Russian Federation NS-2184.2003.2, MK-3507.2004.2,
MK-2203.2005.2, by INTAS grants YSF 04-83-3015, YSF 05-109-5359, by the
grant of the Russian Science Support Foundation and  by contract IAP V/27
of Belgian Science Policy and by IISN.

\end{document}